\documentclass[12pt]{article}

\usepackage{scicite}   

\usepackage{times}

\usepackage{graphicx}
\usepackage{mathrsfs}
\usepackage{amsmath, amssymb}
\usepackage{dsfont}
\usepackage[section]{placeins}

\usepackage{t1enc}
\usepackage[utf8]{inputenc}

\newenvironment{sciabstract}{%
\begin{quote} \bf}
{\end{quote}}

\newcounter{lastnote}

\newcommand{\bra}[1]{\langle#1|}
\newcommand{\ket}[1]{|#1\rangle}

\newcommand{\projNonzero}{\hat M_0^\perp}
\newcommand{\projZero}{\hat M_0}

\newcommand\coinL{\mathord{\leftarrow}}
\newcommand\coinR{\mathord{\rightarrow}}

\usepackage{color}

\title{Probing Measurement Induced Effects in Quantum Walks via Recurrence}

\author{Thomas Nitsche,$^{1,\ast}$ Sonja Barkhofen,$^{1}$ Regina Kruse,$^{1}$ Linda Sansoni,$^{1}$ \\
Martin {\v S}tefa{\v n}\'{a}k,$^{2}$ Aur\'el G\'abris,$^{2}$ V\'aclav Poto\v cek,$^{2}$ Tam\'as Kiss,$^{3}$ \\ Igor Jex$^{2}$ and Christine Silberhorn$^{1}$\\
\normalsize{$^{1}$University of Paderborn, Applied Physics,}\\
\normalsize{Warburger Stra\ss e 100, 33098 Paderborn, Germany}\\
\normalsize{$^{2}$Czech Technical University in Prague,}
\normalsize{Faculty of Nuclear Sciences and Physical Engineering,}\\
\normalsize{B{\v r}ehov\'a 7, 115 19 Praha 1 -- Star\'e M{\v e}sto, Czech Republic}\\
\normalsize{$^{3}$Wigner Research Center for Physics,}\\
\normalsize{Konkoly-Thege M. u. 29-33, H-1121 Budapest, Hungary}\\
\normalsize{$^\ast$To whom correspondence should be addressed: E-mail: tnitsche@mail.upb.de}}

\date{}

\begin{document}




 \maketitle

\begin{sciabstract}
Teaser: The effects of projective measurements on the quantum mechanical evolution of a particle can be simulated with coherent light.\\
Measurements on a quantum particle unavoidably affect its state, since the otherwise unitary evolution of the system is interrupted by a non-unitary projection operation.
In order to probe measurement-induced effects in the state dynamics using a quantum simulator, the challenge is to implement controlled measurements on a small subspace of the system and continue the evolution from the complementary subspace. A powerful platform for versatile quantum evolution is represented by photonic quantum walks due to their high control over all relevant parameters. However, measurement-induced dynamics in such a platform have not yet been realized.
Here we implement controlled measurements in a discrete-time quantum walk based on time-multiplexing. This is achieved by adding a deterministic out-coupling of the optical signal to include measurements constrained to specific positions resulting in the projection of the walker's state on the remaining ones.
With this platform and coherent input light we experimentally simulate measurement-induced single particle quantum dynamics. We demonstrate the difference between dynamics with only a single measurement at the final step and those including measurements during the evolution.
To this aim we study recurrence as a figure of merit, i.e.\ the return probability to the walker's starting position, which is measured in the two cases. We track the development of the return probability over 36 time steps and observe the onset of both recurrent and transient evolution as an effect of the different measurement schemes, a signature which only emerges for quantum systems. Our simulation of the observed one particle conditional quantum dynamics does not require a genuine quantum particle but is demonstrated with coherent light.
\end{sciabstract}

\section{Introduction}

In classical mechanics any observable can be determined at any time without influencing the state of a particle or its future evolution.
In quantum mechanics, contrarily, the measurement has a substantial impact since it interrupts an otherwise unitary evolution of a closed quantum system.
After the measurement the state of the quantum particle is projected onto an eigenstate of the observable according to some probability distribution, resulting in an irreversible change often referred to as the collapse of the wave function. 
This characteristic is at the basis of the quantum mechanical picture and is incorporated among the core postulates of quantum mechanics \cite{dirac_principles_1982,von_neumann_mathematical_1955}.

Since quantum measurements affect the wavefunction of the system, their potential goes beyond being a tool to determine the state of the quantum particle. They can indeed be utilized to influence the evolution of the system implementing a so-called conditional dynamics. This feature of quantum mechanics has found applications such as fault tolerant quantum computation \cite{shor_scheme_1995}, measurement based one-way quantum computation \cite{raussendorf_one-way_2001}, and universal probabilistic quantum gates in linear optical quantum computation \cite{knill_scheme_2001}.

The discrete-time quantum walk (DTQW), the quantum mechanical analogue of the random walk, is a well-established platform for the simulation of particle dynamics \cite{aharonov_quantum_1993}
, and has been used as an instrument to explore the quantum advantage by transferring concepts developed in the classical context to the quantum realm \cite{bouwmeester_optical_1999,perets_realization_2008, peruzzo_quantum_2010,schreiber_photons_2010, broome_discrete_2010,schreiber_decoherence_2011,regensburger_photon_2011,owens_two-photon_2011,sansoni_two-particle_2012,schreiber_2d_2012, di_giuseppe_einstein--podolsky--rosen_2013, poulios_quantum_2014,xue_observation_2014,cardano_quantum_2015,nitsche_quantum_2016,boutari_large_2016,barkhofen_measuring_2017}.
Yet, the border between classical and quantum world is not strictly set and depends on the semantic environment.
A prominent example is coherence, that denotes superpositions, which on one hand is considered to be a fundamental quantum feature, while on the other hand it is the basis for classical optics \cite{streltsov_quantum_2017}.
Indeed, considering a single particle system, it has been demonstrated that using coherent light is sufficient to simulate arbitrary quantum states and unconditional quantum dynamics, e.g.\ transport phenomena \cite{bouwmeester_optical_1999,perets_realization_2008,schreiber_photons_2010,regensburger_photon_2011,schreiber_2d_2012,nitsche_quantum_2016,boutari_large_2016,barkhofen_measuring_2017} or decoherence \cite{broome_discrete_2010,schreiber_decoherence_2011,regensburger_photon_2011}.
These experiments were possible due to the well-established
equivalence of propagation of coherent light across a linear optical network and the dynamics of a single quantum particle (see p. 106 in \cite{paul_introduction_2004}). 
Moreover, coherent states of light can be adopted to simulate certain aspect of quantum measurements along the same principle.

In order to enable a consistent study of the effect of quantum measurements, our present work is centered around two observation schemes: a \textit{reset scheme} with a measurement only at the end of the experiment, and a \textit{continual scheme} where the system is subjected to measurements during the entire course of the evolution.
In classical mechanics the measurements can be made non-invasively, and thus we do not expect any difference in the dynamics of the two schemes.
However, the situation for a quantum system might be radically different.
Young's celebrated double-slit experiment with single particles can be considered as an elementary example: observing the flashes only at the back screen corresponds to the reset scheme, while monitoring which slit a particle has taken corresponds to the continual scheme.
As it is well-known, the presence or absence of the interference pattern depends on this choice \cite{jaeger_two_1995,englert_fringe_1996}.
Moreover, the two schemes are closely related to the two scenarios of the Elitzur--Vaidman bomb-testing problem \cite{elitzur_quantum_1993} in which the bomb is either live or a dud.

An in-depth comparison of the two observation schemes can be made by considering the recurrence problem of a walk, where the central question is whether the walker returns to the origin.
The probability associated to this event is called recurrence probability \cite{polya_uber_1921}.
While for classical random walks the recurrence properties are independent of the observation scheme, for quantum dynamics it greatly depends on it \cite{stefanak_recurrence_2008,grunbaum_recurrence_2013}.

In this paper we experimentally probe the difference between the continual and the reset scheme on a quantum system by simulating the evolution of a quantum walker on a line and study the return to the origin, i.e. the\ recurrence.
The schematic representation of the two situations is depicted in Figs.~1 (a) and 1 (b), respectively.
The walker is simulated by a coherent light pulse and the projection operations for the continual scheme are modeled by controlled absorptive losses that we term \textit{sinks}, since they are acting as traps for the quantum walker.
In contrast to homogeneous losses, e.g.\ passive beam-splitter losses which affect all subspaces identically, controlled losses may be chosen to act only on certain subspaces.
In the simplest scenario the loss may be inflicted at a single position, leaving all the other ones intact.
In case the controlled losses approach unity at the selected sites, they yield a good approximation of a projection onto the subspace corresponding to the complement of the affected subspace.
For example, a sink placed at position zero will realize a projection onto the subspace of non-zero positions. 
Consequently, introducing controlled sinks for a specific subspace while preserving the coherence of the complementing subspace mimics the effect of the non unitary measurement operation on a single quantum particle, which always results in the loss or survival of the particle depending on the result of the projection. 
As such, sinks acting on coherent light allow for the simulation of projective measurements.
With the recurrence probability as a figure of merit we clearly demonstrate the fundamental impact of the measurement operation on a quantum system, resulting in two different recurrence regimes for the two observation schemes: recurrent \cite{stefanak_recurrence_2008} or transient \cite{grunbaum_recurrence_2013}.

\section{Results}

Before turning our attention to the two different notions of recurrence and how they can be adopted in our experimental setup to probe measurement induced effects, we briefly introduce discrete-time random walks and quantum walks.
A random walk is a stochastic process where the walker at every time step hops randomly between neighboring sites according to prescribed rules. The position $x$ of the walker on the one-dimensional lattice after $t$ steps is a random variable with a probability distribution $p(x,t)$. 
In DTQW the walker becomes a quantum mechanical object which after $t$ steps can be in a state of occupying several sites in a coherent superposition, described by a wave function $|\psi(t)\rangle$.
In the so called coined DTQW the evolution of the quantum walker at every time step is described by a unitary operator $\hat{U} = \hat{S}\hat{C}$ according to
\begin{equation}
\label{step:DTQW}
|\psi(t)\rangle = \hat U |\psi(t-1)\rangle.
\end{equation}
The coin is an internal degree of freedom which sets the direction of the next hopping.
The unitary coin operator $\hat{C}$ is applied to this internal degree of freedom every time before the hopping takes place.
The hopping itself is described mathematically by a conditional shift operator $\hat S$, which moves the walker from its previous position $x$ to the nearest neighbors $x\pm 1$ according to the internal coin state.
For more details we refer the reader to the Methods section. 
A common choice of the coin operator for a two-state quantum walk is given by $\hat C_H = \hat{\mathds{1}}\otimes \hat H$, where $\hat H$ is the standard $2\times2$ Hadamard matrix which mimics a fair coin toss \cite{ambainis_one-dimensional_2001}.
In the following we also assume this particular choice, albeit our experiment allows more generic settings for a coin operator.

\subsection{Measurement schemes and recurrence}

In order to demonstrate the fundamental difference between the two observation schemes mentioned above, our choice of a measurement-dependent property is recurrence.
In dynamical systems, recurrence denotes the return of the system to its initial state (or very close to it). 
In certain classes of Hamiltonian systems this is guaranteed by the Poincar\'e recurrence theorem \cite{poincare_sur_1890}, in others, like classical random walks, scenarios in which the particle never returns to its initial position can have nonzero probability. 

We begin with a description of the return probability using random walks as a fundamental model system, considering both the continual and the reset measurement schemes, and then transfer these ideas to quantum walks.

Recurrence in random walks on lattices has been studied by G. P\'olya already in 1921 \cite{polya_uber_1921}. 
In his seminal paper, P\'olya investigated the events of \emph{first return} of the walker to the origin of the walk, i.e.\ to position $x=0$.
Let us denote by $q(0,t)$ the probability that the first return happens in step $t$. 
Events of first return are mutually exclusive and the sum
\begin{equation}
\label{eq:Polya}
{\cal P}(T) = \sum\limits_{t=1}^{T} q(0,t),
\end{equation}
is the probability of recurrence within the first $T$ steps. 
The recurrence probability ${\pmb P} \equiv \lim_{T\rightarrow\infty}{\cal P}(T)$ is called P\'olya number. 
A walk is called recurrent if the walker returns to the origin with certainty, i.e. if its P\'olya number equals unity $({\pmb P}=1)$. 
As shown by P\'olya, unbiased random walks on a line and a plane are recurrent. 
However, already in three-dimensional space the random walker has a non-zero probability of never returning to the starting point, i.e. ${\pmb P}<1$, which is called transient. 

In an experiment, P\'olya's original approach to recurrence of a random walk would correspond to the continual scheme with local measurements only at the origin and the evolution continued after the measurement. 
However, we can also investigate the recurrence of a random walk in an experiment using the reset scheme \cite{stefanak_recurrence_2008}. 
In such a case we consider an ensemble of identical walkers used individually in independent trials with the same settings.
In the $t$-th trial we let the $t$-th walker evolve freely for $t$ steps after which we  observe the origin. 
The probability to detect the walker in the $t$-th trial is given by the  probability at the origin $p(0,t)$.
Since the trials are independent, the product $\prod_t\left(1-p(0,t)\right)$ is the probability that we do not find the walker in position 0 at any $t$. 
The recurrence probability within the first $T$ steps is then defined as the probability of the complementary event, i.e. that we find the walker at the origin in at least one of these $T$ trials, and hence given by the formula
\begin{equation}
\label{eq:resetQW}
{\cal P}_{r}(T) =  1 - \prod\limits_{t=1}^T\left(1-p(0,t)\right).
\end{equation}
A random walk is called recurrent in this observation scheme if ${\pmb P}_{r} \equiv \lim_{T\rightarrow\infty}{\cal P}_{r}(T) = 1$, and transient otherwise.
The property whether a random walk is recurrent turns out to be independent of the observation scheme, i.e.\
\begin{equation}
{\pmb P}=1 \Longleftrightarrow {\pmb P}_{r} = 1,
\label{eq:equiv}
\end{equation}
albeit for transient random walks the actual values of the recurrence probabilities ${\pmb P}$ and ${\pmb P}_{r}$ may differ \cite{stefanak_recurrence_2008-1}. 
This equivalence arises from the fact that the measurement does not change the state of a classical walker, therefore, the dynamics of a random walk in the two schemes remains the same (see also the Supplementary information for a concise summary).
A common explanation for this is that a classical walker has a definite position at any time step, and the probabilistic description stems only from our lack of knowledge.

Let us now consider the same scenarios in a quantum walk.
The recurrence probabilities within the quantum mechanical picture have been previously derived in the literature both for the continual and the reset scheme\cite{stefanak_recurrence_2008,grunbaum_recurrence_2013}.
Interestingly, it has been found that for quantum walks the equivalence in Eq.~\ref{eq:equiv} does not hold between the two notions of recurrence.
Since the governing dynamics remains the same and the only difference between the two cases are the observation schemes, we can attribute the change in the recurrence properties being induced by the measurement alone.
In the reset scheme \cite{stefanak_recurrence_2008-1,stefanak_recurrence_2008}  the quantum walker undergoes uninterrupted unitary evolution according to the operator $\hat U$ (in our case a Hadamard walk) for $t$ steps, after which its state is described by the wave function $|\psi(t)\rangle$.
Performing the measurement at the origin yields the presence of the walker there with probability
\begin{equation}
p(0,t) = |\langle 0|\psi(t)\rangle|^2.
\label{eq:reset_p}
\end{equation}
Identically to the classical case, we use Eq.~\ref{eq:resetQW} to express the recurrence probability in the reset scheme, ${\cal P}_{r}(T)$.
On the other hand, in the continual scheme we have to monitor the presence of the quantum walker at the starting point $x = 0$ after each step.
At each step we examine the result of this measurement and continue the experiment only if the particle is not found, otherwise we 
stop and proceed to the experiment with the next instance.
If the walker is not found at the origin, for the subsequent evolution we describe its state by a wave function obtained after setting the probability amplitude at this particular position to zero while leaving all other positions intact.
This post-selection operation is mathematically described by the projection operator $\projNonzero = \hat{\mathds{1}} - |0\rangle\langle 0|$, which is alternated with the otherwise unitary evolution of a DTQW described by the operator $\hat U$.
We describe the event that the walker is found at the origin by projecting its state onto the position eigenstate $|0\rangle$ by the operator $\projZero = |0\rangle\langle0|$, note however, that the evolution is not continued any further.

Provided that it was not detected in the previous $t-1$ steps, the state of the walker after $t$ steps of the DTQW with the continual observation at the origin is given by the \textit{conditional} wave function
\begin{equation}
|\psi_c(t)\rangle = \frac{1}{\sqrt{s_{t-1}}} \hat{U}\, (\projNonzero\, \hat{U})^{t-1} |\psi (0)\rangle,
\label{eq:condWF}
\end{equation}
prior to any measurement due at step $t$. For convenience, we have introduced $s_{t-1}$ to denote \textit{survival probability} until the step $t$, i.e.\ the probability that the walker has not crossed the origin during the first $t-1$ steps \cite{grunbaum_recurrence_2013}, as
\begin{equation}
s_{t-1} = \left\Vert(\projNonzero\, \hat{U})^{t-1} |\psi (0)\rangle\right\Vert^2.
\label{eq:survprob}
\end{equation}
Besides the physical meaning, the factor also ensures proper normalization of the conditional wave function (Eq. \ref{eq:condWF}).
Using $|\psi_c(t)\rangle$ we determine the {conditional probability} to find the walker at any possible position $x$ after $t$ steps, provided that it was not found at the origin in the previous steps, according to the formula
\begin{equation}
p_c(x,t)
= \left\vert\langle x|\psi_c(t)\rangle\right\vert^2.
\label{eq:condit_p}
\end{equation}

The first return probability after $t$ steps, $q(0,t)$, which is required to determine recurrence in the continual scheme (see Eq.~\ref{eq:Polya}), can be expressed as the product of the survival probability of Eq.~\ref{eq:survprob} and the conditional probability of Eq.~\ref{eq:condit_p}, according to the formula
\begin{equation}
q(0,t) = \left\vert\langle 0|\hat{U}\, (\projNonzero\, \hat{U})^{t-1} |\psi (0)\rangle\right\vert^2 =  s_{t-1} \ p_c(0,t).
\label{eq:continual_q}
\end{equation}
By substituting $q(0,t)$ into Eq.~\ref{eq:Polya} we obtain the return probability ${\cal P}(T)$ for the continual scheme of a quantum walk.

The two schemes described above yield clearly different evolutions for the quantum walker, resulting in fundamentally different recurrence properties, as shown in \cite{grunbaum_recurrence_2013}.
As mentioned above, the relation in Eq. \ref{eq:equiv} does not hold for quantum walks.
The quantum walk on a line in the reset scheme 
 is always recurrent (${\pmb P}_{r}=1$), except for certain trivial coins, while in the continual scheme
it is always transient (${\pmb P}<1$). 
In the particular example of a one-dimensional DTQW with the Hadamard coin the P\'olya number becomes ${\pmb P} = \frac{2}{\pi}$ \cite{ambainis_one-dimensional_2001}.

In the continual scheme the recurrence properties of the walk are determined by the survival probabilities. 
In an ideal experimental implementation, the survival probability corresponds to the success rate of the experiment, i.e.\ to how many walkers are detected in the $t^{\text{th}}$ step when measurements are included, compared to the number of walkers at the initialization.
This establishes the equivalence between the projective measurements, $\{\hat{M}_{x}\}$, and controlled losses acting on a coherent wavefunction, which we implement experimentally by sinks. 
The walkers which encounter a sink at position $x$, and are consequently coupled out of the walk, are the ones that do not {\it survive} the projection operation, thus do not contribute to the survival probability $s_{t-1}$.
In a real experiment, it is crucial to distinguish between losses arising from the action of the measurement, i.e.\ the losses introduced by the sinks, and homogeneous losses which are due to setup imperfections and always occur in any experimental setup.
These homogeneous losses can be compensated while preserving the effect of the sinks.
Thus, we are able to apply the appropriate normalization to the wavefunction while extracting the first return probabilities (Eq. \ref{eq:continual_q}) from the experimental data (more details on how these probabilities are calculated are given in the Methods section).

\subsection{Experimental Implementation}

In order to experimentally demonstrate the distinction between the two measurement regimes, we implement a time-multiplexing quantum walk setup based on a fiber loop, see Fig.~2 \cite{schreiber_photons_2010,nitsche_quantum_2016}.
Here, a coherent laser pulse plays the role of the walker, using polarization as the coin degree of freedom.
The main principle of time-multiplexing is to translate the position degree of freedom into the time domain by splitting up the initial pulse, routing it through fibers of different lengths and thus introducing a well-defined delay between the contributions traveling different paths (see Methods section for details).
The different time-bins of the arrival time histogram of one roundtrip can then be interpreted as the walker's positions $x$ in space.
Here, we extend the capabilities of the architecture by implementing a deterministic in- and out-coupling of the optical signal. This is achieved by adopting two electro-optic modulators (EOMs) in the two fiber arms which are fast enough to switch the polarization at each position individually.
This way we are able to control whether a pulse in a given time bin is fed back into the loop and continues with the quantum walk evolution or is coupled out and routed to the measurement unit.
Such action realizes an absorbing sink at a particular position in the walk and, in addition, allows us to observe how much light has been coupled out.
By programming the switching times of the EOMs, we can easily realize schemes with measurements at $x=0$ in each step (roundtrip),  and schemes with measurements only after a particular number of steps.
Let us note that even when controlled losses were introduced to simulate measurement operations, we were able to observe the evolution over a large number of steps (36 steps). This was made possible by a drastic reduction of the homogeneous losses compared to previously used setups, by optimization of the wavelength, and by upgrading the out-coupling from probabilistic to deterministic. In the present setup, the round trip efficiency exceeds 80\,\% whereas previous results reported efficiencies in the range of 35--50\,\% \cite{schreiber_photons_2010,nitsche_quantum_2016}.

For the purposes of the reset scheme (Fig.~1 (a)), we need to measure the unconditional probability $p(0,t)$ according to Eq.~\ref{eq:reset_p}. 
Technically speaking, this corresponds to a standard implementation of a Hadamard DTQW \cite{schreiber_photons_2010, regensburger_photon_2011, nitsche_quantum_2016}, as the amplitudes of the coherent pulses in the respective time-bins are the experimental representations of the probability amplitudes of the walker's wave function $|\psi(t)\rangle$.
Therefore, the probability distribution $p(x,t)$ can be reconstructed by coupling out all light in a certain step $t$ and determining the relative count rates. 
Afterwards, the experiment is reset and the wavefunction evolves until step $t+1$, when it is coupled out and measured, and so on.
 
In order to probe the recurrence in the continual scheme, it is the probability of the \textit{first} return to the origin $q(0,t)$ that has to be determined.
A sink at the position $x=0$, implemented by deterministically coupling out all light from the corresponding time-bin, realizes the projection operator $\projNonzero$ of the conditional dynamics.
Constantly coupling out all light at the origin before examining a certain step $t$ ensures that the light from the pulse which is detected at the origin in this step has indeed reached it for the first time (see Fig.~1b).

We experimentally measure the evolution of the intensity distribution for a photonic quantum walk on a line with a Hadamard coin from step 0 to step 36 for both schemes and retrieve the unconditional and conditional probability distributions of the walker, shown in Fig.~3.
Subfigures 3 (a) and 3 (b) show the intensity distributions in both cases for step 30.
A comparison of  the two distributions for the overall time-evolution in  Fig.~3 (c) and 3 (d) reveals a signature of the conditional dynamics, implemented by the absorption at the origin, which is manifested in a region of low intensity around $x=0$ in Fig.~3 (d). 
From this data, we extract the conditional and unconditional  probabilities $p_c(0,t)$ and $p(0,t)$ to find the walker at $x=0$ for each step according to equations \ref{eq:condit_p} and \ref{eq:reset_p}, respectively, as well as the survival probability $s_{t-1}$ of Eq.~\ref{eq:survprob}. We then calculate the two different recurrence probabilities according to formulas \ref{eq:Polya} and \ref{eq:resetQW}, and present them in Fig.~4.  Here, the error bars are derived from simulations
where we consider all the systematic inaccuracies of the experiment (deviations in the coin angle, non-perfect polarization transmission of the PBS, errors on the rotation angle of the EOMs) and retrieve the maximum deviation we can expect from the ideal setting parameters (See Methods for details).

\section{Discussion}
The experimental results allow us to study the impact of measurement on the evolution of a quantum system. Our analysis relies on the fact that the evolution of a quantum walker in a one dimensional quantum walk can be either recurrent or transient depending on the choice of the observation scheme. The two kinds of behaviour are clearly demonstrated in Fig.~4 where we have plotted the recurrence probability as a function of the number of steps.
While for the reset observation scheme the recurrence probability ${\cal P}_{r}(T)$ gradually increases over the measured number of steps, for the continual observation scheme ${\cal P}(T)$ quickly starts to saturate as predicted, at $2/\pi$ within the error bars.
Since for a classical walker the two observation schemes would yield identical behaviour, the difference in the two quantum cases is the consequence of the invasiveness of quantum measurement. This fundamental property of quantum mechanics is captured in its essence by the concept of projective measurements, implemented in our experiment by sinks.
Nevertheless, we stress that our experimental results are obtained with coherent light, proving that classical resources are sufficient to simulate single particle quantum dynamics, including the action of projective measurements. 

In conclusion, using the example of recurrence in a DTQW we have experimentally demonstrated the fundamental difference between unitary and measurement-induced dynamics implemented via the reset and the continual observation schemes.
This has been made possible by utilizing a deterministic in- and out-coupling mechanism in a fiber loop based DTQW, which simulates the action of projective measurements in specific positions on the lattice.
The replacement of the probabilistic out-coupling allows maintaining a
high signal to noise ratio throughout the experiment and reaching
almost 40 steps of the simulated quantum walk even in the presence of absorbing sinks.

The realization of controlled sinks significantly enhances the
capabilities of time-multiplexed setup as a quantum simulator,
pointing well beyond the investigation of recurrence in DTQW.  While
in the present experiment we have implemented one sink at the position
$x=0$ constantly over all steps, by reprogramming the signal, we can
realize sinks at any desired positions at any time step, even polarization sensitive.
With the existing coherent light source this allows the investigation of various types of single particle open quantum dynamics.
For example by incorporating two sinks at the edges we can simulate
quantum transport on a finite line or a ring \cite{stefanak_percolation_2016}.
In addition, by altering the coin operator locally at individual
positions \cite{schreiber_decoherence_2011,nitsche_quantum_2016} we
can simulate quantum transport in systems of topological insulators
\cite{kitagawa_exploring_2010} or under influence of point defects \cite{wojcik_trapping_2012}, 
as well as spatio-temporal phase disorder \cite{joye_dynamical_2010}. 
To address situations with multiple particles we have to take the next
experimentally challenging step and use multiple single photon input states.
The presented setup has been designed to be compatible with single
photon sources and therefore is an attractive candidate to perform
experiments such as time-multiplexed boson sampling \cite{motes_scalable_2014},
and multi-walker quantum walks including conditional measurements.

\section{Materials and Methods}
\subsection{Quantum Walk on a Line and Recurrence}
The quantum walker on a line can be in a superposition state of positions $x \in \mathbb{Z}$ and coin states labeled by $c \in \{\coinR,\coinL\}$ given by 
\begin{equation}
| \psi (t)\rangle =  \sum_x \sum_c a_{x,c}(t)| x \rangle \otimes | c\rangle
\end{equation}
with time-dependent amplitudes $a_{x,c}(t) \in \mathbb{C}$.
In analogy to its classical counterpart --- the random walk --- the dynamics of the standard DTQW is given by the alternating application of a coin toss $\hat{C}$ and a conditional shift $\hat{S}$ in space, i.e. a single step is carried out by applying the unitary operator $\hat U = \hat{S}\hat{C}$ according to the rule
\begin{equation}
| \psi (t+1)\rangle = \hat{U}| \psi (t)\rangle= \hat{S}\hat{C}| \psi (t)\rangle.
\end{equation}
The conditional shift operator $\hat S$ moves the walker on the line to the right (left) when its internal coin state is $\ket{\coinR}$ ($\ket{\coinL}$)
\begin{equation}
\hat{S} = \sum_x \left( \ket{x + 1}\!\bra{x}\otimes \ket{\coinR}\!\bra{\coinR} + \ket{x - 1}\!\bra{x}\otimes \ket{\coinL}\!\bra{\coinL}\right)\,.
\end{equation}
In order to achieve a non-trivial evolution of the quantum walker the state of the coin is altered by a toss prior to the conditional shift.
For the coin toss we consider the commonly studied case of the Hadamard operator, which is in the standard basis of the coin space $\{\ket{\coinR} = (1,~0)^T,\ \ket{\coinL} = (0,~1)^T\}$ represented by the matrix

\begin{equation}
\hat{C}_\mathrm{H}
= \hat{\mathds{1}}\otimes\frac{1}{\sqrt{2}}
\begin{pmatrix} 
1 & 1 \\ 
1 & -1 
\end{pmatrix}.
\label{eq:Hadcoin}
\end{equation}

When discussing the recurrence of a DTQW we consider the walker starting from the position $x=0$, i.e. the initial state of the walk reads
\begin{equation}
|\psi(0)\rangle = |0\rangle\otimes|\phi\rangle,
\end{equation}
where $|\phi\rangle$ is some initial state of the coin which can be an arbitrary superposition of the basis states $\ket{\coinR}$ and $\ket{\coinL}$. 
The recurrence probability for a DTQW on a line is independent of $|\phi\rangle$ in both observation schemes \cite{stefanak_recurrence_2008,grunbaum_recurrence_2013}.

\subsection{Experimental setup}

To implement the dynamics in the two observation schemes we use our well-established time-multiplexing quantum walk based on a fiber loop, which has proven to provide great resource efficiency, high homogeneity and long-lasting stability against uncontrolled dephasing.
Additionally it allows for a stable, coherent evolution that can be monitored over a sufficient number of steps \cite{schreiber_photons_2010,nitsche_quantum_2016}. 
The quantum walker is simulated with a weak coherent pulse at a wavelength of 1550 nm with its polarization representing the coin state of the walker. 
The initial pulse is horizontally polarized, corresponding to the state $\ket{\coinR}$.
The step operation is implemented by directing the light with a polarizing beam splitter (PBS 1, see Fig.~2) to two single-mode fibers (SMF) of different length depending on its polarization. 
Consequently, we introduce a well-defined delay between the two polarization components such that each position in each step is represented by a unique arrival time signature. 
Contrary to the previously used setup \cite{schreiber_photons_2010,nitsche_quantum_2016}, 
here the first polarizing beam splitter (PBS) serves for the incoupling into the setup (Port B).
By manipulating the polarization of the pulses with two fast switching electro-optic modulators (EOM) in front of PBS 2, we control whether the pulses are fed back into the loop or are directed to the detection unit. 
Pulses going into port D remain in the loop and undergo the Hadamard coin operation $\hat{C}_\mathrm{H}$ of Eq.~\ref{eq:Hadcoin}, realized by the half-wave plate (HWP).
On the other hand, pulses directed into port C are actively coupled out and directed to the detection unit, which corresponds to the implementation of a sink.
The switching signals of both EOMs exhibit a rise/fall-time of $<$ 5\,ns and can be spaced as close together as 50\,ns, allowing to address the individual position in the walk spaced by 100\,ns (limited by the dead time of the detectors). 
The polarization-resolving detection unit consists of a third PBS and a photon-counting apparatus implemented with two superconducting nanowire single photon detectors (SNSPD) with efficiencies of 60\,\% and\ 70\,\% respectively and a dead time of $\approx$\,100~ns.

A substantial improvement of the active in- and out-coupling over previous setups is the replacement of the probabilistic out-coupling ---formerly realized by partially reflecting mirrors--- which was an additional source of loss in each roundtrip.
The switching accuracy of the EOMs and thus the in- and out-coupling efficiency exceeds 99\,\%.
In combination with optimization of the optical components for a wavelength of 1550\,nm, this greatly improves the roundtrip efficiency from less than 50\,\% in the former setup \cite{nitsche_quantum_2016} to above 80\,\%.
Since roundtrip losses cause an exponential decay of the remaining intensity in the loop, such a significant loss reduction leads to a much higher number of observable steps above the signal-to-noise ratio (i.e. 36 steps even in the presence of sinks in comparison to 28 which was the record in this setup so far \cite{schreiber_decoherence_2011}).
Such a loss-optimized fiber loop with the highly efficient deterministic in- and out-coupling is a well-suited platform for introducing single photon quantum input states as well.

The data presented here is obtained by combining data sets collected from measurement runs in which the quantum walk is implemented for given numbers of steps. At the end of each measurement run all remaining light is coupled out to the detection units, and the results are recorded.
To study the dynamics up to a large number of steps, a power level that is as high as possible would be desirable. Practically, however, since we are using photon-counting detection, high power levels lead to a distortion of the results due to detector saturation.
This problem can be overcome by employing two different initial power levels for measurements. For data sets corresponding to the first 5 steps, we introduce a neutral-density filter (ND filter) with an optical density of 8 in the input beam to guarantee reliable detection within this range. 
For higher numbers of the steps, however, this level of input power yields unsatisfactory results. 
To achieve sufficient visibility at higher numbers of steps, we record data sets employing ND filters with optical densities of 7 and 6.
While these powers cause the detectors to saturate initially, we are able to resolve up to 21 and 36 steps, respectively. Restarting the experiment with a repetition rate of 8 kHz, we use integration times of 10 seconds for steps up to 5, 60 seconds up to step 21 and 3600 seconds for steps 22 to 36.
Evaluation of the results beyond 36 steps is rendered difficult by low signal to noise ratio (see Supplementary Information) and interlacing of time-bins from consecutive round trips.
{This is related to the specific design of the loop: the round trip time is around $2\ \mu$s while the separation between different time bins is around $100$ ns. After 20 steps the arrival times corresponding to different positions in the lattice are spread over the complete round trip time, leading to time bin interlacing for consecutive steps. However the loop is designed such that the round trip time is not an integer multiple of the position separation, and thus after step 20 we have interlacing time bins but no overlap between them. From step 40 the time bins start to overlap and do not represent unique positions in the lattice. Let us note that this limitation can be overcome by using a longer round trip time or shorter position separations, which can be easily done by choosing suitable fiber lengths.}

\subsection{Calculating probabilities from experimental data}

\noindent{\it Reset scheme}. In the photonic quantum walk we experimentally record intensity distributions by measuring count rates at the photodetector.
For the reset scheme we let the light make $t$ round trips in the loop and then measure the number of counts $N(x,t)$ at all possible positions $x$.
The probabilities $p(x,t)$ are calculated by dividing the counts at the respective positions by the overall number of counts in the step under examination.
Doing so we normalize out the effect of the unavoidable homogeneous losses introduced by all passive optical components in the loop.
In particular, the probability at the origin at time $t$ is given by the ratio
\begin{equation}
p(0, t)=\frac{N(0,t)}{\sum\limits_{y} N(y,t)}.
\label{p(0,t)}
\end{equation}
{\it Continual scheme}. In the continual scheme we want to obtain experimentally the first return probability after $t$ steps $q(0,t)$.
To ensure that the pulses detected after $t$ steps have not crossed the origin before we implement sinks at $0$ at previous times. 
Coupling out all light afterwards we record count rates $N_c(x,t)$ at every possible position $x$. 
Normalizing by the total number of counts at step $t$ we obtain the 
conditional probability 
$$
p_c(x,t)=\frac{N_c(x,t)}{\sum\limits_{y} N_c(y,t)}.
$$
To obtain the first return probability $q(0,t)$ according to Eq. \ref{eq:continual_q} we have to multiply $p_c(0,t)$ by the probability $s_{t-1}$ that the walker has not been absorbed during the first $t-1$ steps and survives until step $t$.
Since we can assume the same homogeneous losses in both experiments (with and without sinks), the survival probability can be calculated as
$$
s_{t-1}=\frac{\sum\limits_{y} N_c(y,t)}{\sum\limits_{z} N(z,t)}.
$$
This leads to the expression for the first return probability
\begin{equation}
q(0,t)=s_{t-1}\ p_c(0,t)=\frac{N_c(0,t)}{\sum\limits_{y} N(y,t)}.
\label{cond:0}
\end{equation}
Thus, assuming only homogeneous losses, the proper first return probabilities can be obtained by dividing the counts at position zero from the continual scheme by the overall counts in the reset scheme.
Let us note that alternatively, the results from the reset scheme can be used to estimate the homogeneous loss rate, and the total counts (photons detected in the ``sink'' and at the final round trip) in the continual scheme used for normalization.
We have analyzed the data both ways and obtained the same first return probabilities up to errors that were markedly smaller than the error bars (Fig.~4).
This supports our homogeneous loss assumption and that these losses are the same in the experiments on both observation schemes.

\subsection{Error bars}
We have identified four sources of systematic errors in our experimental setup, cf.\ also \cite{nitsche_quantum_2016}: first, the detector and power dependent detection efficiencies, which were determined in a separate measurement; 
second, the different losses experienced in different paths due to dissimilar coupling efficiencies and path geometries, which are estimated in an independent measurement with an accuracy of $\pm 1\,\%$; 
third, the exact angle of the coin HWP which can only be determined up to an error of $\pm 0.15^\circ$;
fourth, the switching accuracy of the EOMs which results in a possible residual transmission of the sinks of 1\,\%.
As input parameters for our simulations we consider the measured values and the estimated error ranges.
We take as reference the output distribution simulated with the measured parameters.
We then run simulations considering all the possible combinations of parameters with maximal error. We finally take as error bar the largest deviation of these simulations from the reference output.
Errors coming from Poissonian statistics, scaling with the square root of the number of click events, are also evaluated however they are a minor contribution compared to the systematic deviation and as such do not contribute significantly to the error bars in Fig.~4.

\bibliographystyle{Science}

\section{Acknowledgments}

The group at Paderborn acknowledges financial support from the Gottfried Wilhelm Leibniz-Preis (grant number SI1115/3-1), from the European Union’s Horizon 2020 research  and  innovation  program  under  the  QUCHIP project no.\ 641039 and from European Commission with the ERC project QuPoPCoRN (no.\ 725366). The group at Prague acknowledges financial support by the Ministry of Education, Youth and Sport (Czech Republic) under grant RVO 14000. M.~\v S., A.~G.\ and I.~J.\ have been partially supported by the Czech Science foundation (GA{\v C}R) project number 17-00844S. T.~K.\ was supported by the National Research Development and Innovation Office of Hungary (Project Nos. 2017-1.2.1-NKP-2017-00001 and K124351).

Author Contributions:
T.N. and S.B. designed, built and conducted the experiments. 
T.N. and S.B. wrote the data analysis and data acquisition software. 
R.K., L.S., M.S., A.G., V.P. and T.K. analysed and interpreted the processed data. 
I.J. and C.S. supervised and gave conceptual advice.
All authors discussed the results and wrote the manuscript.

Competing interests: The authors declare no competing interests.

All data needed to evaluate the conclusions in the paper are present in the paper and the Supplementary Materials. Additional data related to this paper may be requested from the authors.

\newpage

\section{Figures and tables}

\begin{figure}[h!tb]
	\includegraphics[width=\columnwidth]{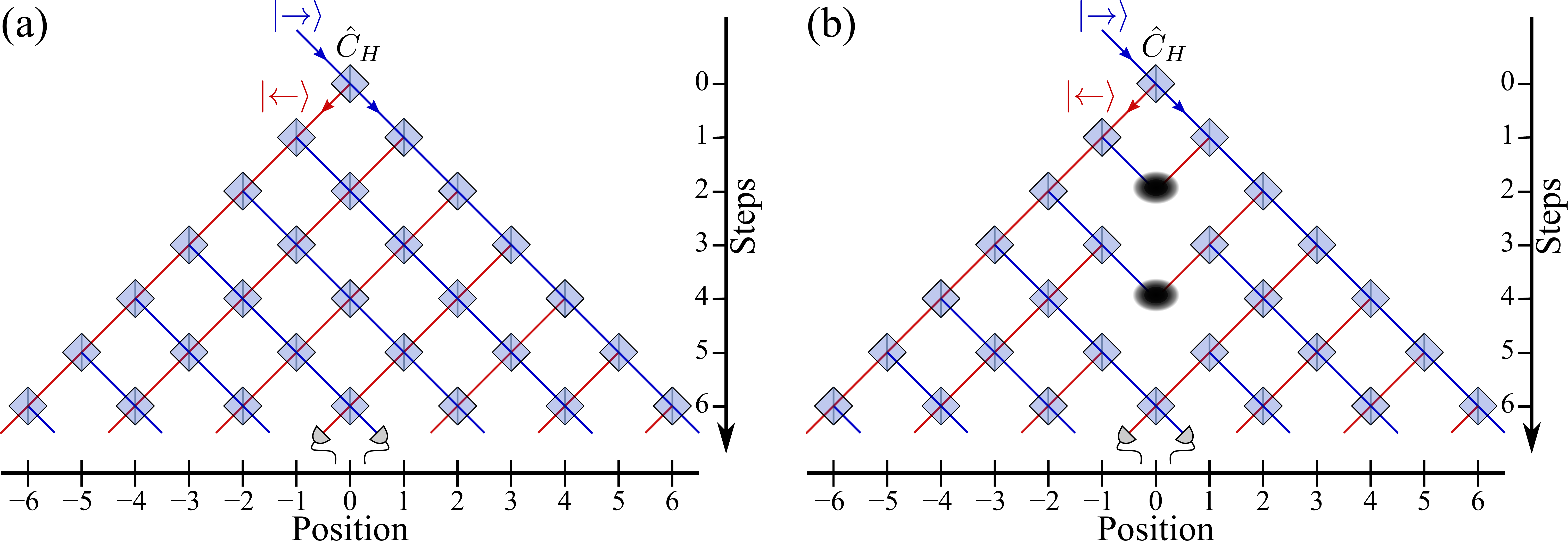}
	\caption{Schemes illustrating the quantum walk implementation of the reset scheme (a) and the continual scheme with the sinks at $x=0$, denoted by black holes (b), both exemplary for a measurement in step 6. Each grey diamond corresponds to the application of the Hadamard coin, followed by the spatial shift. 
	}
	\label{fig:schemes}
\end{figure}

\begin{figure}[h!tb]
	\centering
	\includegraphics[width=0.9\columnwidth]{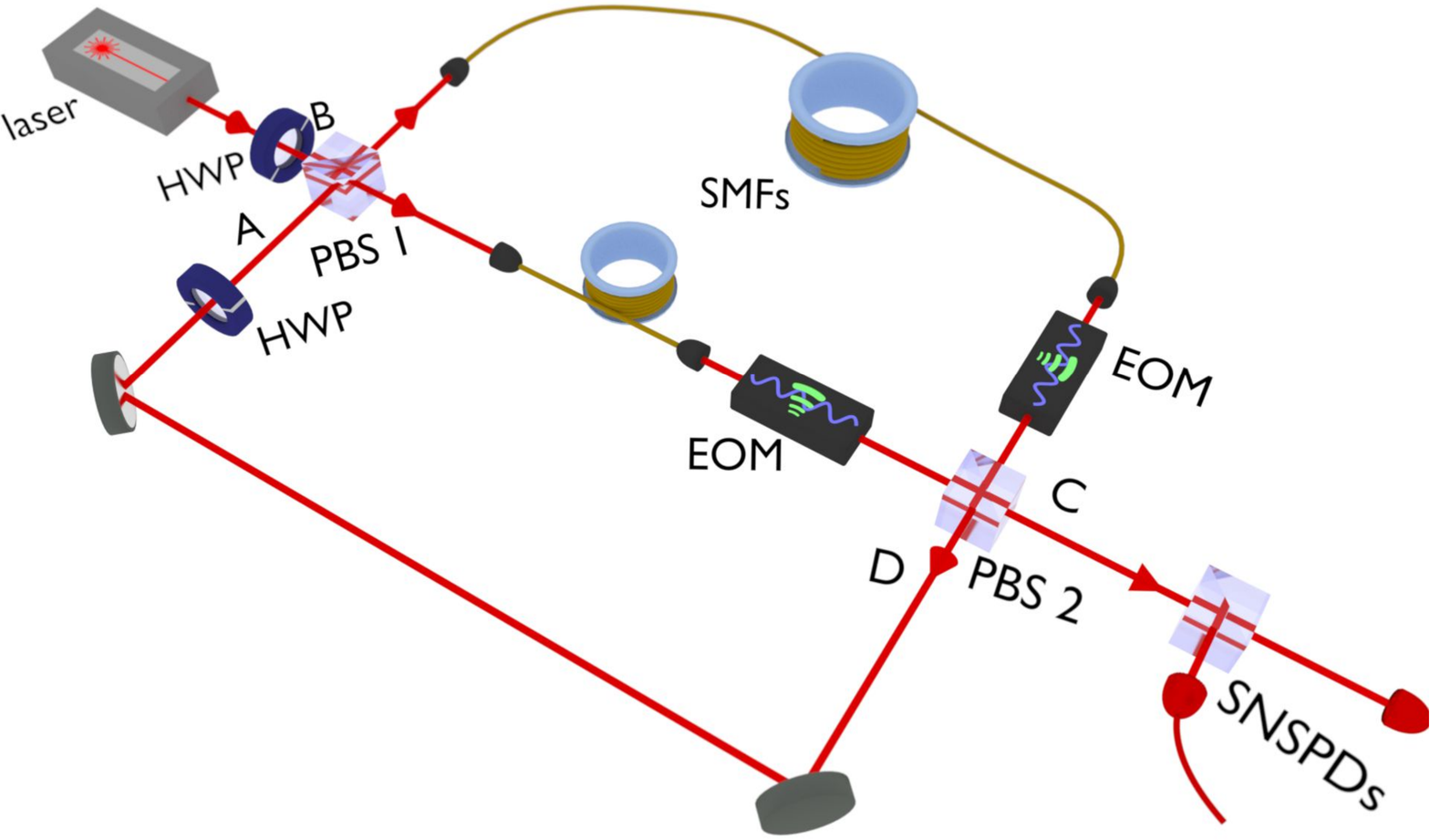}
	\caption{Schematic of the experimental setup of the time-multiplexed quantum walk with active in- and out-coupling realized by two electro-optic modulators (EOMs), see Methods section for details. The active control of 	the switches allow us to implement in the time domain both the continual and reset schemes, physically equivalent to the spatial representations in Figure~1 in one setup.}
	\label{fig:setup}
\end{figure}
\newpage

\begin{figure}[h!tb]
	\includegraphics[width=\columnwidth]{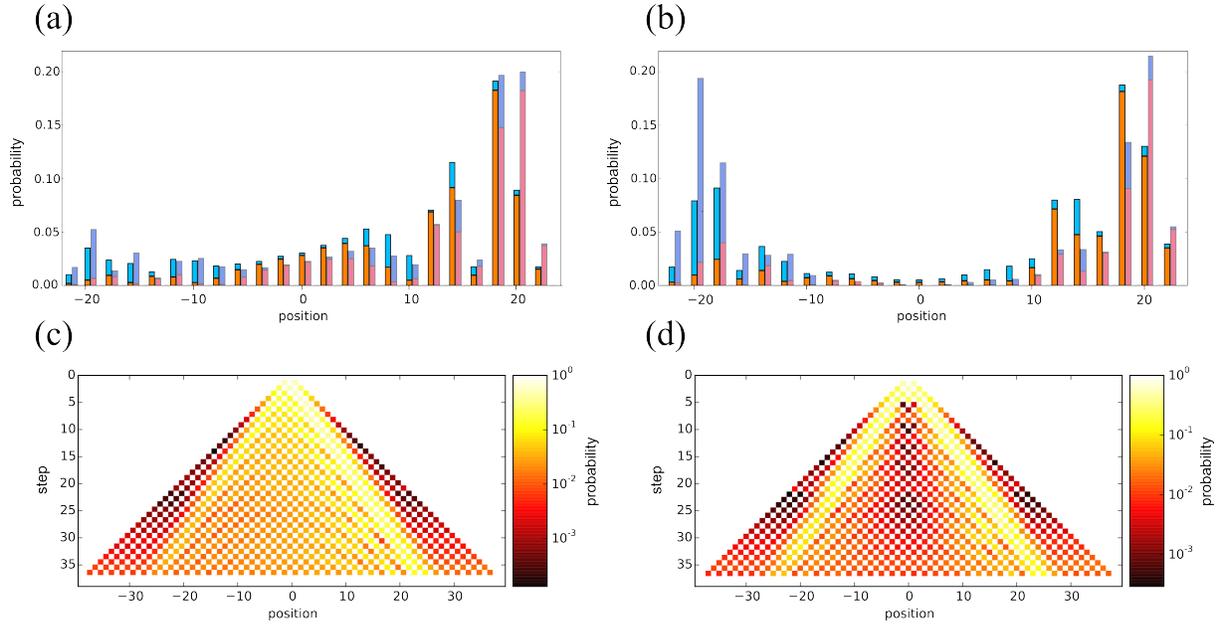}
	\caption{Upper half: Upper half: Intensity distribution in step 30 for the reset scheme (a) and the continual scheme (b). Horizontally-polarized light is represented by the orange (experimental data) and the red (numerical data) bar charts, while vertically-polarized light is depicted by the light blue (exp. data) and the dark blue (num. data) bar charts. The errors bars are omitted for clarity (for error analysis see Methods Section).
		Lower half: Evolution of the experimentally observed intensity distribution over the positions as the walk evolves from step 0 to step 36 in a chessboard diagram with a logarithmic color scale: (c) shows the unitary evolution free of measurements for the reset scheme and  (d) conditional evolution with the sinks for the continual scheme. Note that in this last case we expect a symmetric distribution with respect to the origin and the remaining asymmetry is due to experimental imperfections in the coin realization and coupling efficiencies for the two polarizations (for a detailed analysis of the experimental inaccuracies see Methods section).}
	\label{fig:active_out}
\end{figure}
\newpage

\begin{figure}[h!tb]
	\includegraphics[width=0.9\textwidth]{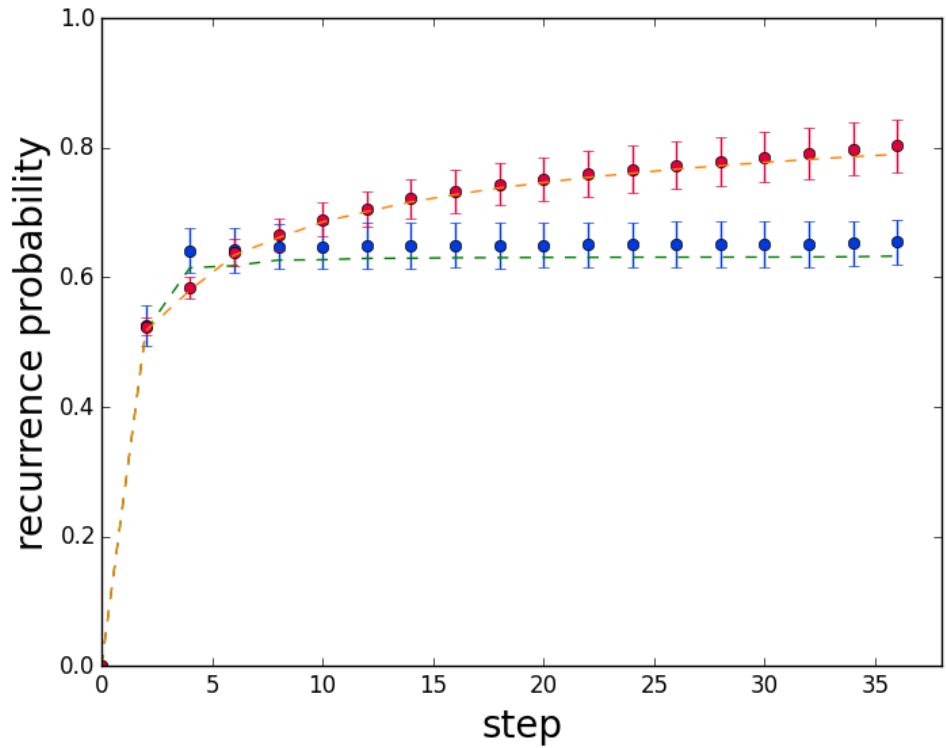}
	\caption{The experimental results for recurrence probabilities in the reset scheme ${\cal P}_{r}(T)$ (red symbols) and the continual scheme ${\cal P}(T)$ (blue symbols).
		The dashed lines give the numerical values that are to be expected from a numerical simulation of the experiment. The overall deviation between experimental and numerical values in the continual scheme is mainly defined by the deviation in step 4, as the contributions of later steps to the sum are small in comparison to the first 4 steps. For the error analysis please refer to Methods.}
	\label{fig:polya}
\end{figure}
\newpage

\section*{Supplementary material}

\subsection*{Recurrence in classical random walks}

We dedicate this section to providing a concise summary on recurrence of classical random walks on infinite lattices.
We start with the approach considered by P\'olya  \cite{polya_uber_1921} who investigated recurrence by analyzing the first return probabilities $q(0,t)$. 
In our terminology, this corresponds to the continual scheme where the starting point of the walk is searched for the presence of the walker after each step. 
The probability that the walker ever returns to the starting point of the walk (P\'olya number ${\pmb P}$) can be written in the form
\begin{equation}
\label{pn:rw:q}
{\pmb P} = \lim_{T\rightarrow \infty} \sum\limits_{t=1}^T q(0,t).
\end{equation}
However, the P\'olya number Eq. \ref{pn:rw:q} can be expressed in terms of the probability of finding the particle at the origin after $t$ steps of the walk $p(0,t)$, where we do not impose additional assumptions on the past of the walker. Indeed, in random walks there is a simple relation between the first return probability $q(0,t)$ and the probability at the origin $p(0,t)$, which leads to an alternative expression for the P\'olya number in the form \cite{hughes_random_1995}
\begin{equation}
\label{pn:rw:p}
{\pmb P} = 1 - \frac{1}{\sum\limits_{t=1}^\infty p(0,t)}.
\end{equation}
This expression is useful for determining the recurrence nature of a random walk. Indeed, ${\pmb P}$ equals 1 if and only if the series  $\sum\limits_{t=1}^\infty p(0,t)$ is divergent, which in turn depends on the asymptotic behavior of $p(0,t)$ as $t$ tends to infinity. For unbiased random walks on $d$-dimensional integer lattices we find that $p(0,t)\propto t^{-\frac{d}{2}}$, which implies that random walks on a line $(d=1)$ and a plane $(d=2)$ are recurrent $({\pmb P} = 1)$, while for $d\geq 3$ the random walks are transient $({\pmb P} < 1)$.

If we investigate the recurrence of a random walk in the reset scheme then the recurrence probability is given by a different formula, namely
\begin{equation}
\label{sup:p:reset}
{\pmb P_r} = 1 - \prod\limits_{t=1}^\infty (1-p(0,t)).
\end{equation}
Nevertheless, it is straightforward to show \cite{stefanak_recurrence_2008-1} that the product in Eq.~\ref{sup:p:reset} vanishes if and only if the series in Eq. \ref{pn:rw:p} is divergent. Hence, for random walks the equivalence 
$$
{\pmb P} = 1 \Longleftrightarrow {\pmb P_r} = 1,
$$
holds. However, the equivalence is broken for quantum walks, as was shown theoretically in \cite{grunbaum_recurrence_2013} and as we have demonstrated experimentally in the present paper. Indeed, due to the influence of measurements on the evolution of a quantum system the relation between the first return probability $q(0,t)$ and the probability at the origin $p(0,t)$ does not hold for DTQWs. Hence, when evaluating the recurrence probability of a DTQW in the continual scheme the Eq.~\ref{pn:rw:p} does not apply and we have to use the Eq.~\ref{pn:rw:q}.

\subsection*{Signal to noise ratio}
In the following section we will discuss the influence of the signal to noise ratio (SNR) on our experimental data which is the limiting factor to the maximum number of steps.
Since every real detector suffers from dark counts we have to subtract them from the measured signal. In the analysis we have considered uniformly distributed noise which we have evaluated in an independent measurement. However, for the last steps the interlacing of time bins increases the noise background due to after-pulsing of the detectors. Since the shape of the afterpulses is broad and they have a fat tail, they may leak into the time window of the consecutive pulse and alter the background. This background is highly correlated with the peak height of the previous pulse. When interlaced pulses occur in between the pulses under consideration for a particular step, their afterpulsing can affect the noise background in a non trivial way.
We were able to correctly extract the signal counts up to step 36 even in the presence of the mentioned noise floor, while from step 38 this fails. To confirm this we have analyzed the SNR and plotted the values of it for each step as shown in Fig. \ref{SNR}.
\renewcommand{\thefigure}{SI.\arabic{figure}}
\setcounter{figure}{0}
\begin{figure}[h!tb]
	\center
	\includegraphics[width=0.8\columnwidth]{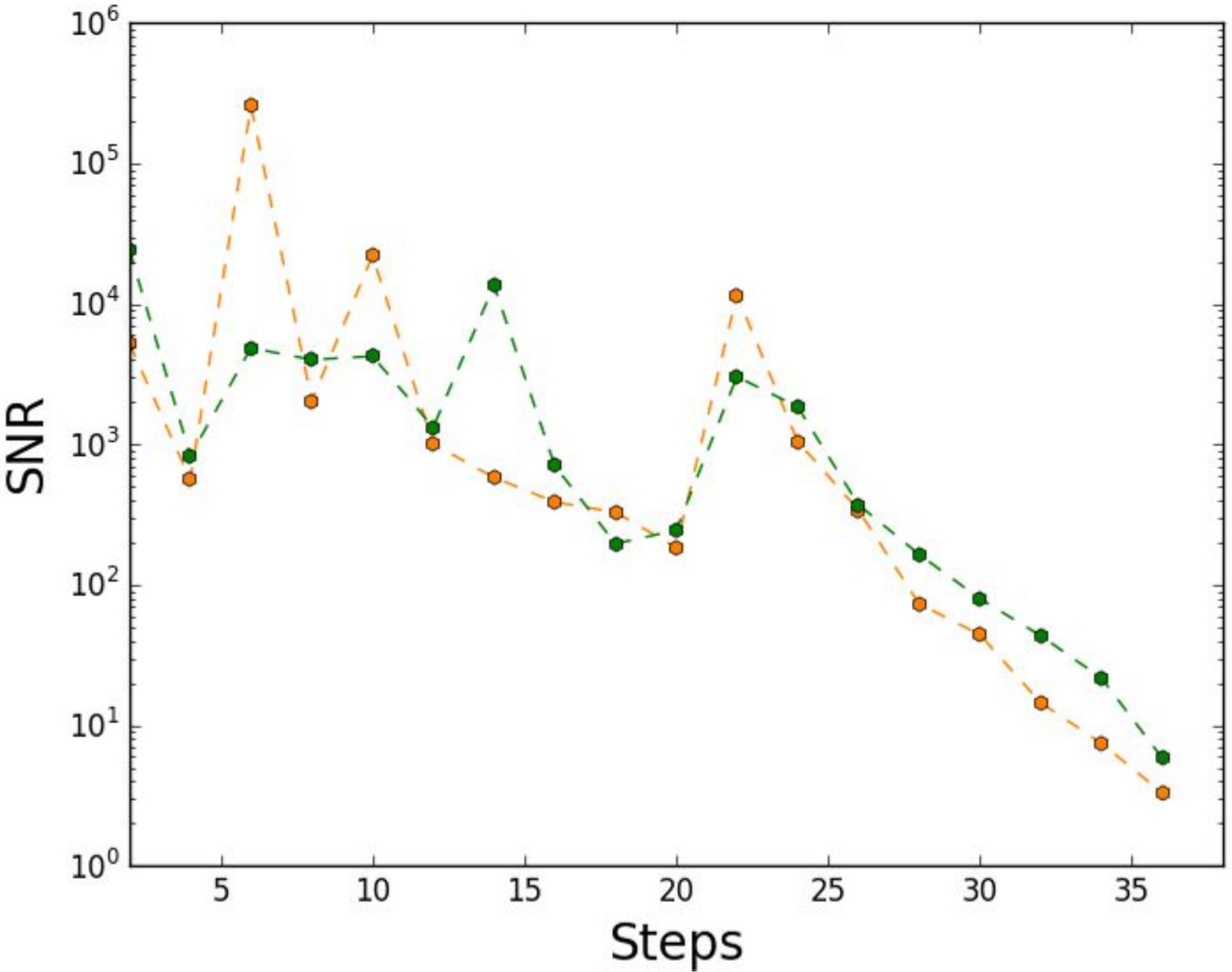}
	\caption{Signal to noise ratio of the experimental data. The green symbols refer to the SNR of the reset scheme measurements, while the orange ones correspond to the continual scheme.	}
	\label{SNR}
\end{figure}

To calculate the SNR we first sum up the overall counts of a pulse train belonging to a specific step taken in 4.8 ns windows centred around the expected arrival time of each pulse. This sum is divided by the noise contributions in windows of the same length close, but not overlapping with the signal windows, which contains a comparable strong influence of the afterpulsing.  We observe that until step 36 the SNR is sufficiently high for reliable data extraction.

\end{document}